\documentclass[10pt,aps,twocolumn,longbibliography]{revtex4-2}
\usepackage{amsmath,graphicx,subfigure,bm,xcolor,mathtools,newtxtext,soul}
\usepackage[ bookmarks=true, colorlinks, linkcolor=blue, urlcolor=blue, citecolor=blue, plainpages=false, pdfpagelabels, final, breaklinks=true ]{hyperref}
\usepackage{ulem}

\newcommand{\calL}[0]{\mathcal{L}}
\newcommand{\bcdot}[0]{\boldsymbol{\cdot}}

\begin{document}
\title{Graphene Josephson Junctions for Engineering Motional Quanta}


\author{Zhen-Yang Peng}
\email[]{pengzhenyang@sjtu.edu.cn}
\affiliation{Wilczek Quantum Center, School of Physics and Astronomy, Shanghai Jiao Tong University, Shanghai 200240, China}

\author{Mehdi Abdi}
\email[]{mehabdi@gmail.com}
\affiliation{Wilczek Quantum Center, School of Physics and Astronomy, Shanghai Jiao Tong University, Shanghai 200240, China}

\begin{abstract}
We propose a hybrid quantum device based on the graphene Josephson junctions, where the vibrational degrees of freedom of a graphene membrane couple to the superconducting circuits. The flexural mode-controlled tunneling of the Cooper pairs introduces a strong and tunable coupling even at the zero-point fluctuations level. By employing this interaction, we show that a parametric process can be efficiently implemented. We then investigate foundational and technological applications of our hybrid device empowered by nonlinear interactions, with fast generation of non-classical mechanical states, and critically enhanced quantum sensing under suitable quantum control. Our work provides the possibility of employing the graphene motional degree of freedom for quantum information processing in circuit quantum nanomechanical structures.
\end{abstract}


\maketitle

\section{Introduction}

Nanomechanical systems as solid-state-based devices have been widely used for studying various quantum behaviors~\cite{nanomech_review_2011, mechanicalsystem_review_2012, Xiang_RMP_2013, nanomech_RMP_2022}. This rapidly growing field of research offers new candidate platforms for investigating fundamental quantum physics~\cite{Arndt_2014, Liao_2016, Abdi_prl_2016, Khosla_PRX_2018}. Also, it has potential for applications in quantum technologies~\cite{Nanomech_nature_2013_transducer, Abdi_2021, Rossi_2018, Rossi_2019, nature_2019_mechosc_sensing, Wang_2024, Rahman_2025}. The quantum control of nanomechanical structures requires its coupling to other quantum devices, which have been proposed and realized in various ways~\cite{Rabl_2010, PRL_2024_spinqubit, Abdi_2019, PRX_2023_QD, Cleland_nphys_2013, nphys_om_review_2022, nphys_hybridqs_review_2020, Abdi_2015}. Based on the hybridization with motional degrees of freedom (DOF) of mechanical objects, quantum nanomechanical systems can be designed with higher flexibility, small dissipations, strong nonlinearity, and ultra-strong couplings~\cite{Chang_RMP_2018, RMP_2019_ultrastrongcouple}. As a consequence, hybrid quantum nanomechanical systems have become a rapidly developing paradigm for studying quantum foundations with technological potential. 

In recent years, the maturing technologies for superconducting circuits have significantly improved their hybridization with nanomechanical systems, and consequently, their manipulation~\cite{Cleland_science_2023, Cleland_NC_2025}.
The graphene Josephson junctions (GJJ) are one of the well-studied and fabricated structures in superconducting circuits~\cite{Beenakker_PRB_2006, Heersche_nature_2007, Julien_natnano_2022_PA, Haller_PRR_2022_GJJSQUID, Kroll_NC_2018, Steel_NC_2018, Lee_2020, Sarker_natnano_2022_PA}.
Such systems provide new viewpoints to explore exotic quantum behaviors in condensed matter systems, such as Andreev bound states detection~\cite{nphys_2017_Andreevstate, Zhang_PRB_2023, Lee_PRL_2024}, helical edge modes~\cite{Rout_NC_2024_helical}, magic-angle twisted graphene control~\cite{Natnano_2022_magicangle, Zheng_PRR_2024_magicangle} and other non-equilibrium dynamics~\cite{Tsumura_2016_APL, PRR_2024_selfheat}. 


In this paper, we propose a hybrid nanomechanical scheme based on GJJ where the junction is a free-standing graphene membrane and thus has a motional DOF. We show that one can engineer the coupling between the flexural modes of the graphene and the superconducting circuit that involves the GJJ at the zero-point fluctuations (ZPF) level. This coupling strength is highly tunable even in the deep strong coupling regime. By numerical analysis we show that such coupling allows for the cooling of the motional modes close to their groundstate via a parametric cooling process. We then investigate and discuss the cat states generation and quantum metrological aspects of our scheme for frequency estimation of the vibrational mode. The possibility of criticality-enhanced quantum sensing has been explored by investigating the QFI of the vibrational mode when close to the critical point. 

This paper is organized as follows: in Sec.~\ref{sec:model} we introduced our hybrid model, in which the ZPF of the motional DOF has been considered in the GJJ. In Sec.~\ref{sec:cooling}, the parametric cooling for the mitional DOF has been analyzed. Then, two potential applications of our model in mechanical quantum information processing, with which the fast generation of mechanical cat states and critical quantum sensing, have been discussed in Sec.~\ref{sec:catstate} and Sec.~\ref{sec:sensing}, respectively. In Sec.~\ref{sec:conclusion}, we summarized our findings. Some technical details are given in the Appendices.

\section{The model}\label{sec:model}

The fundamental setup of our proposed hybrid quantum device is based on a superconducting circuit and a ballistic graphene layer. The monolayer graphene has been demonstrated as Josephson junctions ~\cite{Beenakker_PRB_2006, Heersche_nature_2007}. The system Hamiltonian is
\begin{equation}
    H_0 = 4E_c ( n- n_{g} )^2 + H_{\text{GJJ}},
\end{equation}
where $E_c$ is the charging energy for the junction, $n$ indicates the number of Cooper pairs across the junctions, and $n_g$ is the offset charge induced by the gate voltage. The Josephson energy in the short-junction regime is given by $H_{\text{GJJ}}= -\Delta_0 \sum_{n=0}^{\infty} \sqrt{1-\tau_n \sin^2{\phi/2}}$(see Appendix.~\ref{ap:Andreev} for more details) with $\Delta_0$ the excitation gap for the superconductor islands, $\phi$ the superconducting phase for the GJJ, and $\tau_n$ the transmission probability for each Andreev mode excited inside the graphene~\cite{Beenakker_PRB_2006}. The latter depends on the chemical potential $\mu$, which in turn can be tuned experimentally, see e.g. Ref.~\cite{Haller_PRR_2022_GJJSQUID}.

Generally, the superconducting excitation gap $\Delta_0$ is large, thus,  the ratio $\Delta_0/E_c \gg 1$ is easily satisfied.
The dynamics of the Andreev bound modes are dominated by the Josephson potential energy part $H_{\text{GJJ}}$, while the phase difference between the two superconductors is small. Hence, the influence of offset charge $n_g$ can be neglected, the perturbative expansion for $\phi$ leads to an anharmonic oscillator form
\begin{equation}
    H_0 = 4 E_c n^2 + \frac{1}{2} \tilde{E}_J \phi^2 - \eta \phi^4,
\end{equation}
where we have introduced
\begin{subequations}
    \begin{align}
        \tilde{E}_J &= \Delta_0\sum_n \tau_n/4,\\
        \eta& = \tfrac{1}{24}\Delta_0 \sum_n (4\tau_n-3\tau^2_n)/16.
    \end{align}
\end{subequations}
In terms of creation ($a^\dag$) and annihilation ($a$) operators, the perturbed Hamiltonian reads ($\hbar=1$)~\cite{Koch_PRA_2007}
\begin{equation}
    H_0 = \omega_r a^{\dagger}a - \tilde{\eta}(a+a^{\dagger})^4,
\label{eq:H_transmon}
\end{equation}
with the resonance frequency $\omega_r{=}\sqrt{8 E_c \tilde{E}_J}$ and the nonlinearity factor $\tilde{\eta}{=}2E_c \eta/\tilde{E}_J$.


\begin{figure}
\centering
\includegraphics[width=\columnwidth]{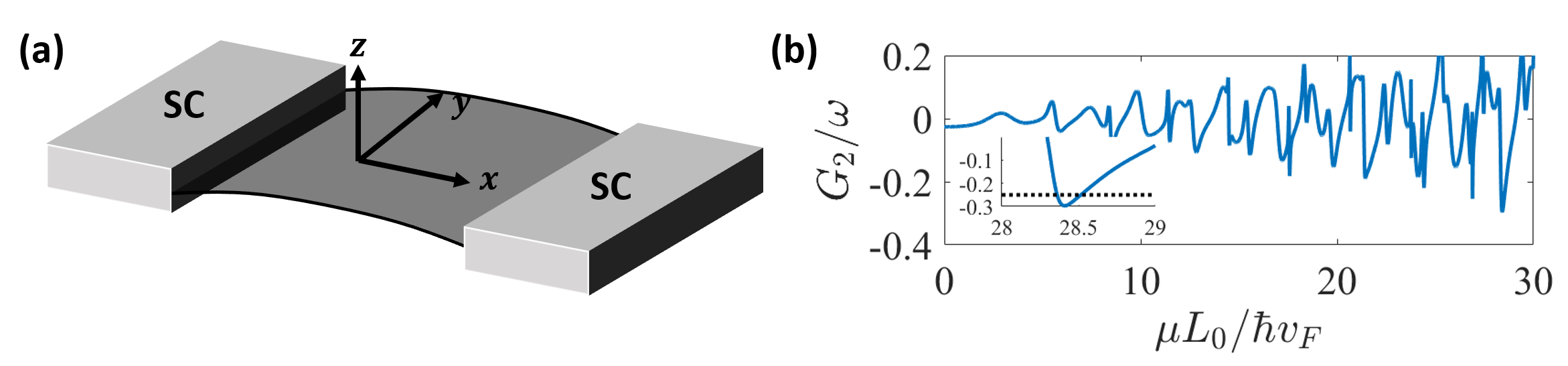}
\caption{(a) Sketch of a Graphene Josephson junction with zero-point fluctuations perpendicular to the static plane. (b) The quadratic coupling strength as a function of the graphene chemical potential. The inset shows that for suitable $\mu$ the effective coupling can even surpass the critical strength (the dashed line).}\label{fig:sketch}
\end{figure}

The Hamiltonian~Eq.~\eqref{eq:H_transmon} is obtained by neglecting the motional DOF of the graphene junction. However, the graphene contact between two superconductors in principle vibrates. To find the coupling between the vibrational DOF of the graphene membrane and the superconducting circuit, the dynamics of ZPF should be considered. Due to the tiny mass and high flexural strength of graphene, the surface tension of graphene is influenced by the ZPF, therefore, vibration perpendicular to the graphene surface ($z$-axis) changes its length, see Fig~\ref{fig:sketch} for an illustration.
Consequently, the Hamiltonian \eqref{eq:H_transmon} becomes a function of $z$, thus, can be Taylor expanded as a series of $z$, which indicates the vibrational DOF of graphene perpendicular to the equilibrium plane ($z=0$).
The Hamiltonian has the following form
\begin{equation}
    H_0(z) = H_0(0) + z \frac{\partial H_0}{\partial z} + \frac{1}{2} z^2 \frac{\partial^2 H_0}{\partial z^2} +\cdots.
\label{eq: Htot_expand}
\end{equation}

The graphene membrane vibrations can be simplified as a set of flexural harmonic oscillations around its equilibrium position. In principle, the oscillation includes all the flexural modes. However, given the typically high mechanical quality factors these modes are well-resolved~\cite{Chen_2013, Will_2017}. Therefore, we only consider the fundamental mode with the lowest oscillating frequency in our study whose profile has the form $\zeta(x) {=} z \cos{(\pi x/L_0)}$, where $L_0$ is the equilibrium length of the junction. The total bent junction length is $L(z){=}2L_0 E(-z^2 \pi^2/L^2_0)/\pi$. Here $E(x)$ is the complete elliptic integral of the second kind. From this, one easily finds that the first and second order perturbation terms in Eq.~(\ref{eq: Htot_expand}) are $\partial H_0/\partial z {=} 0$ and $\partial^2 H_0/\partial z^2 {=} (\pi^2/2L_0)\partial H/\partial L$, respectively. Therefore, the hybrid system Hamiltonian including both circuit and vibrational degrees of freedom and the coupling between them reads

\begin{align}
        H' &= \omega_r a^{\dagger}a -\tilde{\eta}(a+a^{\dagger})^4 + \omega b^{\dagger}b \nonumber\\
        &+ G_2(a+a^{\dagger})^2(b+b^{\dagger})^2,
\label{eq:H_fin}
\end{align}
where the 4th- and higher-order coupling terms has been neglected because of $|\partial \tilde{\eta}/\partial L| \ll |\partial \tilde{E}_J/\partial L|$.
Here, the vibrational mode with frequency $\omega$ and effective mass $M$ is quantized as $z{=}z_{\text{zpf}}(b+b^{\dagger})$ with $z_{\text{zpf}}{=}\sqrt{\hbar/2M \omega}$ the ZPF for the vibration.
The coupling rate is given by
\begin{equation}
G_2 = \frac{\pi^2 \Delta_0}{32L_0}\sqrt{\frac{2E_c}{\tilde{E}_J}}\sum_{n=0}^{\infty} \frac{\partial \tau_n}{\partial L} z^2_{\text{zpf}}.
\label{eq:G2}
\end{equation}
This can be controlled by tuning the charging energy $E_c$, and the transmission probability $\tau_n$ (chemical potential $\mu$). The flexible adjustment of the nonlinear interaction between circuit and vibrational modes offers promising applications in designing appropriate quantum dynamics and generating non-classical states for this hybrid quantum device. 

\section{Parametric Cooling and Effective Mechanical Nonlinearity}\label{sec:cooling}

The quantum control for the vibrational mode requires the graphene membrane to be near its quantum ground state. Unless the oscillating frequency $\omega$ is high enough to ensure $k_B T {\ll} \hbar \omega$, the cooling of the vibrational mode is needed.
The "parametric cooling" process for the vibrational mode can be engineered via coherently driving the superconducting circuit, $H_{\text{dri}}=A(ae^{\text{i}\omega_D t} + \text{H.c.})$, where $A$ is the drive amplitude and $\omega_D$ is its frequency. To understand the cooling process, we move to the rotating frame at the drive frequency $U(t)=e^{\text{i}\omega_D a^{\dagger}a t}$, the total Hamiltonian has the form
\begin{equation}
    \begin{split}
    H_{\text{tot}} &= \delta a^{\dagger}a -6\tilde{\eta}(a^{\dagger}a)^2 + \omega b^{\dagger}b + 2G_2 a^{\dagger}a(b+b^{\dagger})^2\\
    & + G_2(b+b^{\dagger})^2 + A(a+a^{\dagger}), \label{eq:H_dri}
    \end{split}
\end{equation}
where the detuning $\delta=\omega_r-\omega_D-6\tilde{\eta}$, and the counter-rotating terms for circuit mode are neglected. We then add the displacement transformation $D(\alpha)$ to $H_{\text{tot}}$, in the weak drive regime we have
\begin{equation}
    \begin{split}
        H^{\prime}_{\text{tot}}& = \omega_a a^{\dagger}a - 6\tilde{\eta}(a^{\dagger}a)^2 + \omega b^{\dagger}b + \Lambda(b+b^{\dagger})^2 \\
        &+ 2G_2 a^{\dagger}a(b+b^{\dagger})^2 - 2\alpha G_2 (a+a^{\dagger})(b+b^{\dagger})^2,
    \end{split}\label{eq:H_para}
\end{equation}
where $\omega_a=\delta-24\alpha^2 \tilde{\eta}$ is the effective circuit frequency, $\omega_b = \omega + 2\Lambda$ is the effective vibrational mode frequency, $\Lambda = G_2 (1+2\alpha)$, and $\alpha=A/\delta$ is the displacement. The last term indicates a parametric interaction, which provides an effective parametric cooling for the vibrational mode under the resonant condition $\omega_a=2\omega_b$.

\begin{figure}
    \centering
    \includegraphics[width=\columnwidth]{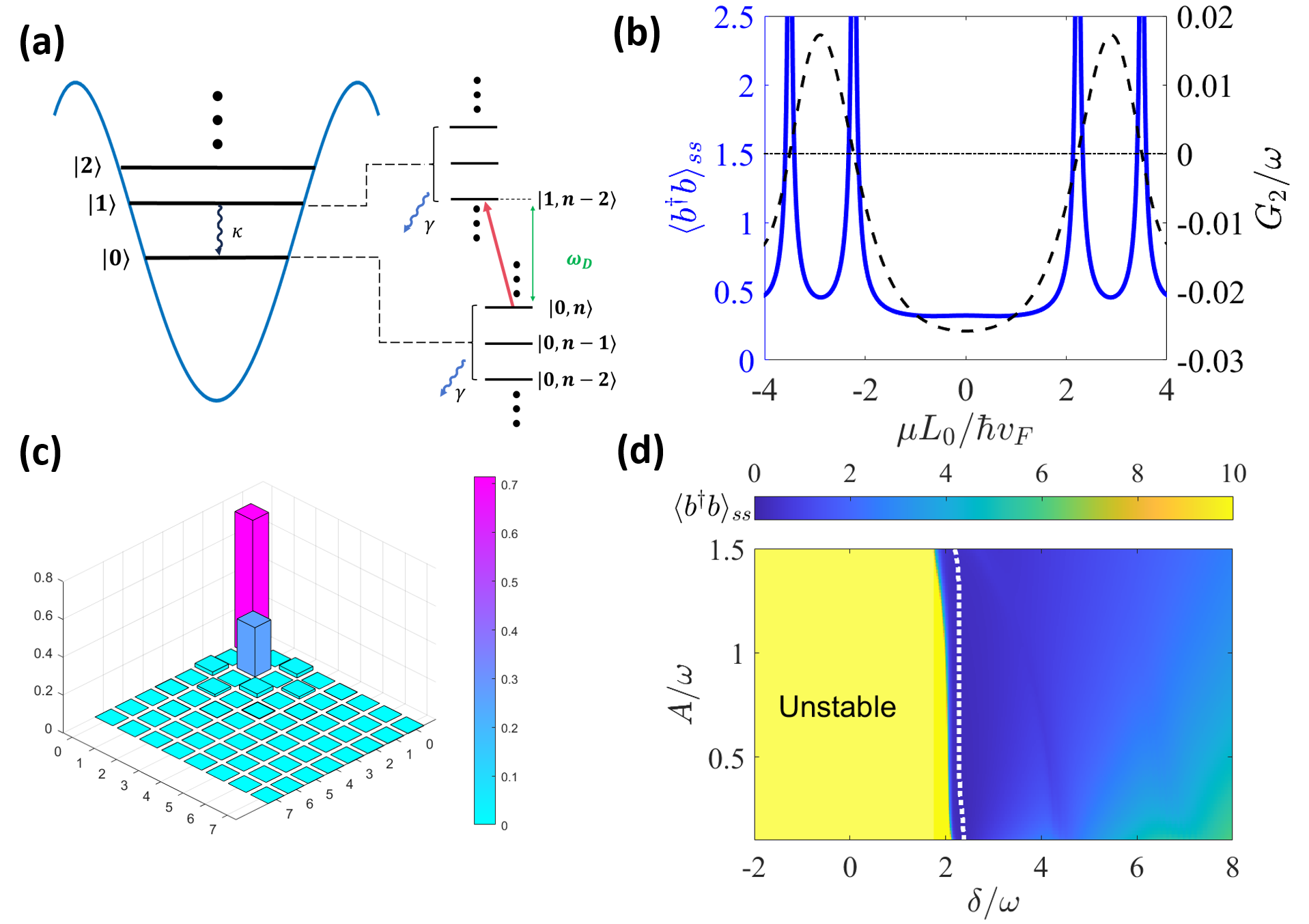}
    \caption{(a) An energy level diagram for the parametric cooling process. (b) Steady-state occupation number for vibrational mode and the corresponding coupling strength $G_2$ for optimal detuning. (c) Absolute value of the lowest elements of the vibrational mode steady-state density matrix after the parametric cooling process. (d) Steady-state occupations for different detuning and coherent drive strength for $\mu=0$. The white dashed-dot line indicates the minimal occupation under different $A$.}
    \label{fig:cooling}
\end{figure}

We prove the efficiency of this cooling scheme by employing the quantum optical master equation
\begin{equation}
	\dot \rho = \calL_0[\rho] +\calL_1[\rho],
\label{eq:master}
\end{equation}
with the free Liouvillian $\calL_0[\bcdot]=-i[H^{\prime}_{\text{tot}},\bcdot]$. The dissipative part of the dynamics is composed of the circuit decay and the damping of vibrational mode $\calL_1=\calL_a +\calL_b$, where $\calL_a[\bcdot]= \kappa D_a[\bcdot]$ and $\calL_b[\bcdot]=\gamma(\bar{n}_b {+}1) D_b[\bcdot] +\gamma\bar{n}_b D_{b^\dag}[\bcdot]$. Here $\kappa$ is the decay rate of the circuit, $\gamma=\omega/Q$ is the damping rate with the mechanical quality factor $Q$, $\bar{n}_b$ is the occupation number of the mechanical bath, and $D_O[\bcdot]{=}O\bcdot O^\dag -(O^\dag O \bcdot +\bcdot O^\dag O)/2$ is the Lindblad superoperator. 


The effective cooling result is illustrated in Fig.~\ref{fig:cooling}, where the system parameters have been considered as $E_c{=}10^6 \ \text{Hz}$, $\hbar \Delta_0 {=} 0.2 \ \text{meV}$, $L_0{=}35 \ \text{nm}$, $v_F{=}2.5\times 10^6 \text{m/s}$, $\omega{=}1 \ \text{MHz}$, the thermal occupation $\bar{n}_b=10$, the decay rate of circuit $\kappa=\omega/10$, the coherent drive strength $A{=}\omega$, and the vibrational mode quality factor is $Q=10^{6}$. Once the $G_2\neq 0$, the steady-state occupation of the vibrational mode is suppressed close to the ground-state level (see Fig.~\ref{fig:cooling}(b)). However, because of the two-phonon nature of the interaction, the vibrational mode can only be cooled down to a mixture of ground and first excited states (shown in Fig.~\ref{fig:cooling}(c)). Meanwhile, the frequency mismatch destroys the cooling and may drive the system into an unstable regime, as shown in Fig.~\ref{fig:cooling}(d).

With the coherent drive, the parametric interaction enables an effective nonlinearity for the vibrational mode. To see this, we apply the Schrieffer-Wolf transformation $\tilde{H}^{\prime}_{\text{tot}}\rightarrow e^{-S} H^{\prime}_{\text{tot}} e^{S}$ to block diagonalize Hamiltonian~(\ref{eq:H_para}) in the limit of $\omega_a {\gg} \omega_b$. We denote
\begin{equation}
    H^{\prime}_{\text{tot}} = H_0 - \lambda V,
\end{equation}
\begin{equation}
\begin{split}
    H_0 &= \omega_a a^{\dagger}a - 6\tilde{\eta}(a^{\dagger}a)^2 + \omega b^{\dagger}b \\
    &+ \Lambda(b+b^{\dagger})^2 + 2G_2 a^{\dagger}a(b+b^{\dagger})^2,
\end{split}
\end{equation}
\begin{equation}
    V = (a+a^{\dagger})(b+b^{\dagger})^2,
\end{equation}
where $\lambda = 2\alpha G_2$, and the generator has been chosen as $S=\lambda S_1$ and $S_1 = (a^{\dagger}-a)(b+b^{\dagger})^2/\omega_a$, which leads to $[H_0,S_1]=V$ for $\omega_a \gg \omega, \tilde{\eta}, \Lambda$. Thus, the transformed Hamiltonian to the leading order becomes
\begin{equation}
    \tilde{H}^{\prime}_{\text{tot}} \simeq H_0 - \xi(b+b^{\dagger})^4. \label{eq:Hamiltonian_SW}
\end{equation}

Since the above Hamiltonian is block-diagonal, the total system can be projected into the subspace where the circuit degree of freedom is in the ground state. Thus, we obtain the final effective Hamiltonian 
\begin{equation}
    H_{\text{eff}} = \omega b^{\dagger}b + \Lambda(b+b^{\dagger})^2 - \xi(b+b^{\dagger})^4,\label{eq:effHamiltonian_b}
\end{equation}
where $\xi {=}  4\alpha^2 G^2_2/\omega_a$. The above Hamiltonian indeed suggests potential applications for our proposed setup, with flexible control and manipulation of the vibrational mode. 

\section{Fast Generation of Cat States}\label{sec:catstate}

\begin{figure}
    \centering
    \includegraphics[width=0.75\columnwidth]{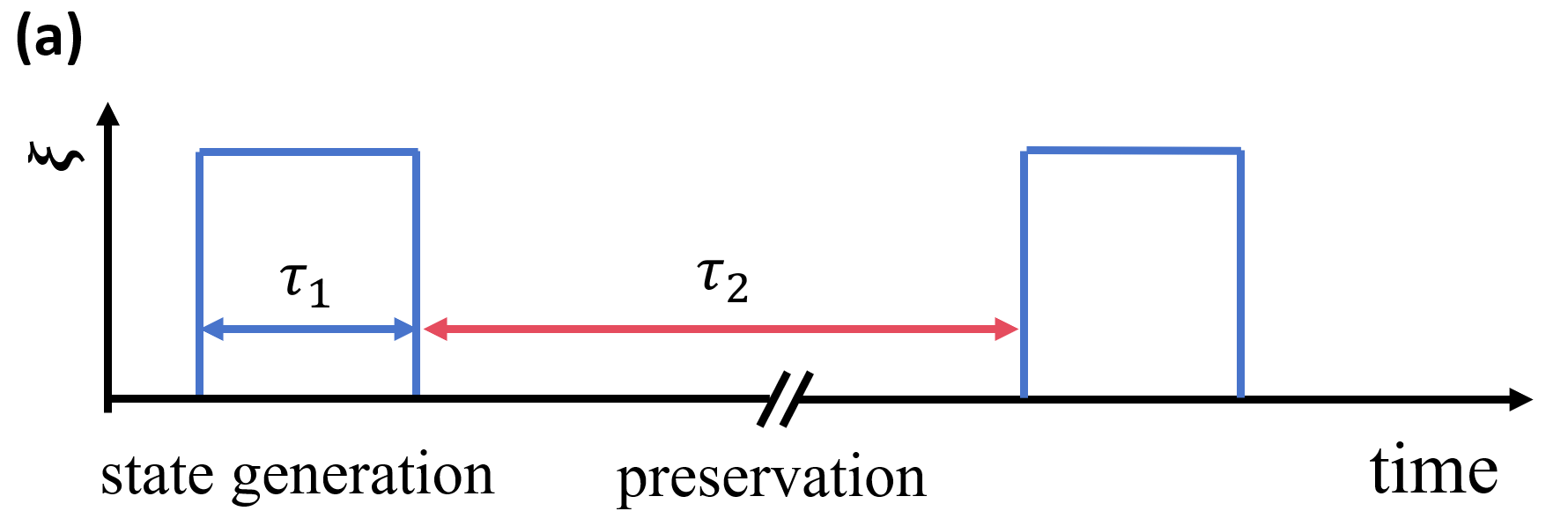}
    \includegraphics[width=0.9\columnwidth]{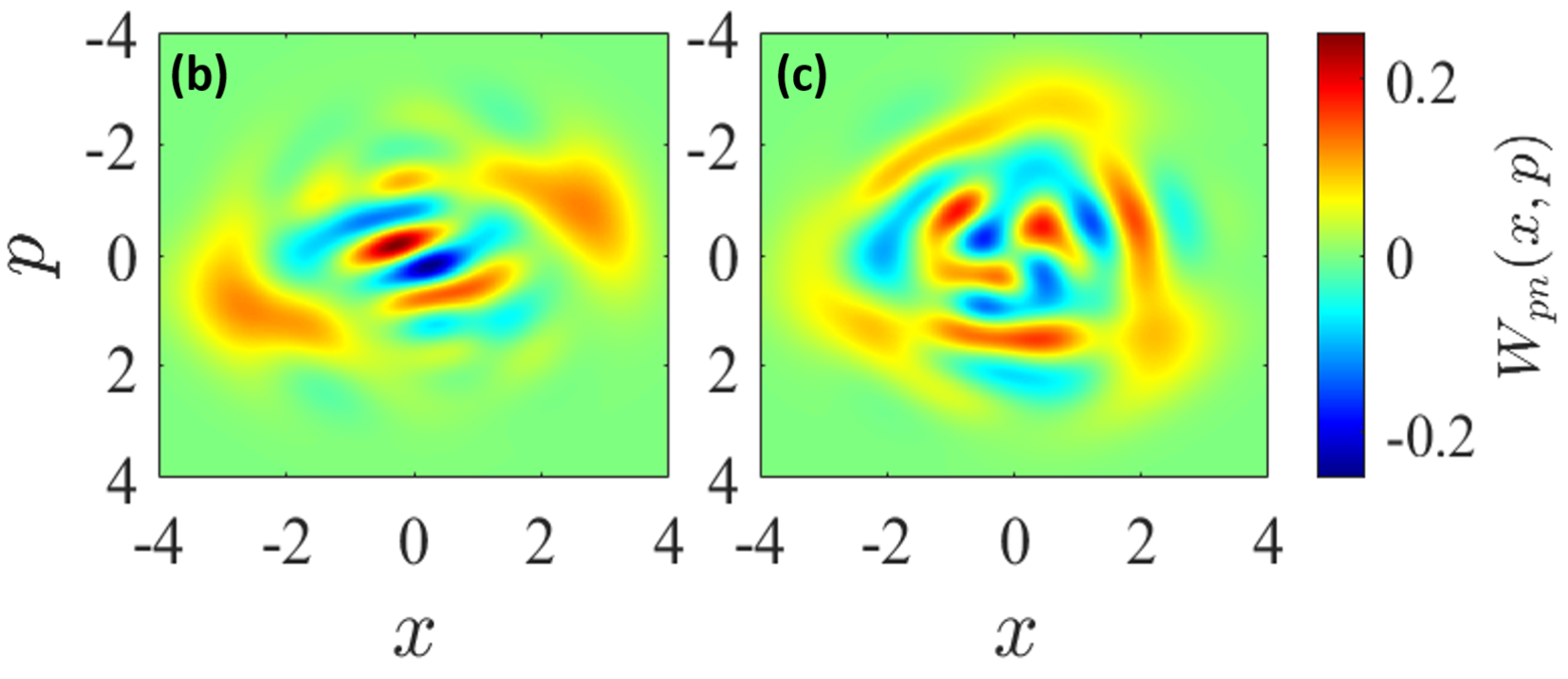}
    \caption{(a) The control sequences for the $m-$ component cat states generation and preservation. Without loss of generality, this sequence can be repeated, as illustrated for the second state generation pulse. (b-c) The Wigner function for the vibrational mode after $\tau_1=\tau_0/2, \tau_0/4$, which generates the 2- and 4-component cat state, respectively. The parameters are chosen as $\xi=10^{-4}\omega$, $\Lambda=0.8\Lambda_c$.} 
    \label{fig:catstate}
\end{figure}

The effective nonlinearity of the effective Hamiltonian~(\ref{eq:effHamiltonian_b}) suggests the potential application in generating multi-component mechanical cat states that can be useful both in quantum foundations~\cite{PRA_2024_catstate}, and quantum technologies~\cite{Devoret_nat_2020_catqubit, He_natcomm_2023, PRXQuantum_2023_catqubit}. After the squeezing transformation $b\rightarrow S(r)b S^{\dagger}(r)$ the effective Hamiltonian has the form
\begin{equation}
    H_{\text{eff}} = \Delta b^{\dagger}b - \xi e^{4r}(b+b^{\dagger})^4, \label{eq:effHamiltonian_sqz}
\end{equation}
where $\Delta{=}\sqrt{\omega^2+4\omega \Lambda}$ is the effective frequency, and the squeezing parameter $\tanh 2r=-2\Lambda/(\omega+2\Lambda)$. The effective system~(\ref{eq:effHamiltonian_sqz}) indeed is critical at $\Lambda_c{=}-\omega/4$. Under the rotating wave approximation, the nonlinear term can be approximated as $H_{\text{Kerr}} = - \tilde{\xi} b^{\dagger 2}b^2$, where $\tilde{\xi}{=}6\xi e^{4r}$ is the Kerr nonlinearity strength. Therefore, once the vibrational mode is prepared in the coherent state $|\beta\rangle$, the $m$-component cat states($m=$2,3,4,...) can be generated under the nonlinear evolution time $\tau_1=\tau_0/m$, with $\tau_0=\pi/2\tilde{\xi}$, see Fig.~\ref{fig:catstate}. 
Compared to Kerr nonlinearities achieved through nonlinear crystals or superconducting transmon circuits, in our case, due to the squeezing amplification, the effective Kerr nonlinearity $\tilde{\xi}$ has an exponential enhancement when $\Lambda \rightarrow \Lambda_c$. Meanwhile, the nonlinear strength $\xi$ can be tuned to a finite value that $\xi \gg \gamma$. Therefore, the effective Kerr nonlinearity strength can be realized in the regime where $\tilde{\xi} \gg \gamma$, ensuring a significantly faster generation time $\tau_1 \ll \gamma^{-1}$. 
After successfully generating the cat states, the nonlinear interaction can be suppressed to $\xi {\rightarrow}0$ for the non-driven case ($\alpha=0$) or equivalently the large circuit frequency limit ($\omega_a{\gg}G_2$), allowing the cat states to evolve only under a small dissipation rate $\propto \gamma$. Hence, the preservation time $\tau_2 < \gamma^{-1}$ ensures sufficient time for implementing further control and manipulation sequences for the corresponding $m-$components cat states. 

In addition, one can explore passive quantum sensing~\cite{Alushi_2024_PRL_metrology} for cat states in the preservation period. After generating the desired states, the coupling $G_2$ can be turned off, allowing the noisy evolution for the cat states only under the free term $\omega b^{\dagger}b$. In this case, the initial state is given by $S^{\dagger}(r)\rho_{\text{cat}}S(r)$, and the cat state acquires the displacement and phase shift for the noisy evolution.

\section{Critical Quantum Sensing}\label{sec:sensing}

As one of the major applications of quantum devices, quantum metrology~\cite{Natphotonic_2011_metrology, Toth_2014_review_qfi, Liu_2020_review_qfi} discusses the estimation of unknown parameters and sets the ultimate precision limit based on the quantum resources for a given quantum system. The upper bound for the estimator that the minimum error can be attained is set by the quantum Fisher information (QFI) $F_o$, which determines the information of the unknown parameter $o$ encoded in a given state $\rho_{o}$. The QFI for the density matrix $\rho=\sum_j \lambda_j |\psi_j \rangle \langle \psi_j|$ through the logarithmic derivative is defined as $F_o {=} 2\sum_{j,k}|\langle \psi_j|\partial_o \rho |\psi_k \rangle|^2/(\lambda_j + \lambda_k)$~\cite{Paris_2009, Toth_2014_review_qfi, Liu_2020_review_qfi}.

\subsection{Zero Temperature}

\begin{figure}
    \centering
    \includegraphics[width=\columnwidth]{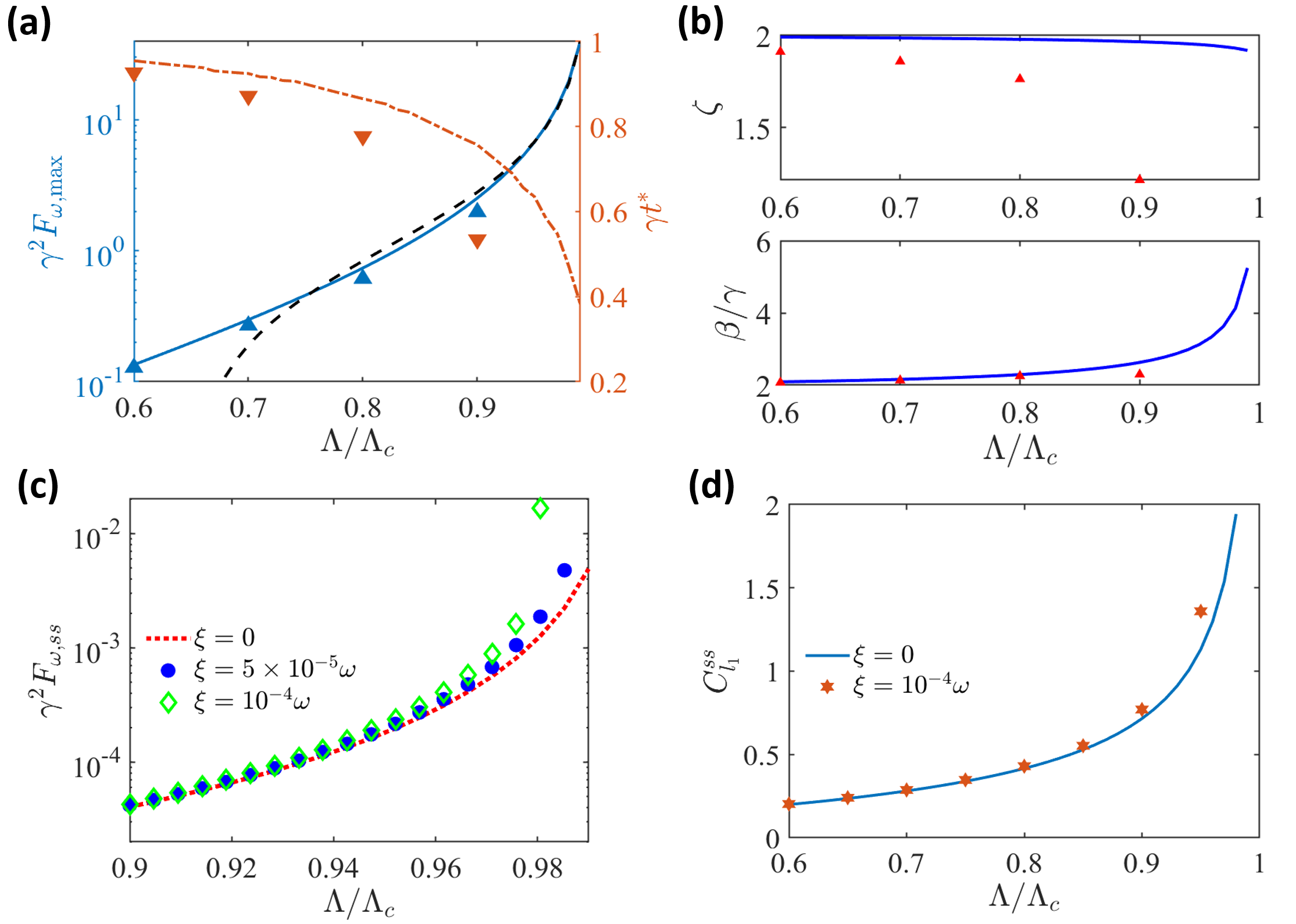}
    \caption{QFI for zero temperature. (a) The maximal QFI (solid blue line) and the corresponding optimal time $t^*$ (dotted red line) for different $\Lambda$. The lines indicate the $\xi=0$ while the triangles correspond to $\xi=10^{-4}\omega$. The dashed black line is the function $(\Lambda_c -\Lambda)^{-1}$ with a proper fitting factor. (b) The fitted parameters $\zeta$ and $\beta$ under zero temperature. The blue lines indicate $\xi=0$ while the red markers correspond to $\xi=10^{-4}\omega$. (c) Steady state QFI vs $\Lambda$ under different $\xi$. (d) The steady state intuitive $l_1$ norm of coherence under computational basis $\{|n\rangle\}$.}
    \label{fig:sensing}
\end{figure}

In our setup, for the case $\xi\rightarrow 0$, the critical coupling $\Lambda_c$ for the Hamiltonian~(\ref{eq:effHamiltonian_b}) indicates the critical point for a second-order quantum phase transition, see e.g.~\cite{Hwang_prl_2015_QRM, Garbe_2022_QST}. This is indeed the operational point that enables the vibrational mode essential for critical quantum sensing.~\cite{Montenegro2024, Sarkar2024, Alushi2025}.
Due to the open quantum systems nature of our system, the QFI can be fitted by the function $F_\omega(t) = C t^{\zeta} e^{-\beta t}$, where the fitting parameters $C$, $\zeta$, and $\beta$ depend on the proximity of $\Lambda$ to the critical point~\cite{gribben2024boundarytimecrystalsac}.
For a closed system, one expects a scaling $\zeta=2$ for the dynamical QFI, while for the open quantum system case we find $\zeta \lesssim 2$ and it begins to decline exponentially because of the dissipations. Therefore, a finite maximum sensitivity $F_{\omega,\max}$ is attainable at the optimal time $t^*=\zeta/\beta$ for frequency sensing. The maximal QFI occurs earlier once the system is closer to the critical point because of the critical amplification of the effective dissipation. Nevertheless, the factor $C$ gets a significant enhancement due to the criticality such that $F_{\omega,\max}$ eventually experiences a power-law growth of $F_{\omega,\max} {\propto} (\Lambda_c-\Lambda)^{-1}$ as $\Lambda \rightarrow \Lambda_c^-$, see Fig.~\ref{fig:sensing}(a).
A similar behavior is observable for the steady-state. For the zero-temperature case, the steady state for the effective Hamiltonian~(\ref{eq:effHamiltonian_b}) is the squeezed vacuum state, hence, the stationary QFI is found as $F_{\omega,ss}=\Lambda^2/32(\Lambda^2_c - \Lambda \Lambda_c)^2$, which is shown as the dotted red line in Fig.~\ref{fig:sensing}(c).

Next we discuss the finite $\xi$ case. Due to the enhanced nonlinearity the environment-induced transitions largely depend on the eigenstates. Moreover, when $\Lambda\rightarrow \Lambda_c$ for finite $\xi$, the eigenstates cannot be perturbatively expanded under the Fock states of the original harmonic oscillator. Therefore, the decoherence of the effective system cannot be described via the conventional local dissipators, see Ref.~\cite{Abdi_prl_2016}.
The master equation describing the effective dynamics has the same form as Eq.~\eqref{eq:master} with $\calL_0[\rho]=-i[H_{\rm eff},\rho]$, while the dissipation part reads
\begin{equation}
    \calL_1[\rho] {=} \sum_{k>j}\Gamma^{jk}\Big\{(\bar{n}_{kj}{+}1)D_{|j\rangle_s\! \langle k |}[\rho] {+}\bar{n}_{kj}D_{|k \rangle_s\! \langle j|}[\rho]\Big\},
\label{eq:effmastereq}
\end{equation}
where $|k\rangle_s$ are the eigenstates of Hamiltonian \eqref{eq:effHamiltonian_b} such that $H_{\text{eff}}|k\rangle_s {=} E_k |k\rangle_s$.
Here, $\Gamma^{jk} {=} \gamma |{}_{s}\langle j|b|k\rangle_s|^2$ is the dissipation rate that describes transition between states $|j\rangle_s$ and $|k\rangle_s$ with $\Delta_{kj} {=}E_k {-} E_j$ and $\bar{n}_{kj}$ is the corresponding bosonic occupation. Detailed derivations are in the Appendix~\ref{ap:effmastereq}.

The results for QFI are plotted in Figs.~\ref{fig:sensing}(a) and (c) as markers. Even though the maximum dynamical QFI does not show appreciable enhancement due to the nonlinearity, the suppression for the scaling $\zeta$ for the finite $\xi$ case (Fig.~\ref{fig:sensing}(b)) leads to shorter $t^*$.
Meanwhile, the negative anharmonicity significantly enhances the steady state QFI.
This is the main difference between our effective model~(\ref{eq:effHamiltonian_b}) and the result obtained for the fully-connected model~\cite{Garbe_2022_QST, Chu_prl_2021, Mishra2021}, which the quartic anharmonic term increases the energy gaps~\cite{Hwang_prl_2015_QRM}, providing a resistance to the quantum criticality. In other words, the influence of higher order terms are equal to the same quantity in the thermodynamic limit, but with a correction to the critical point, see Ref.~\cite{Garbe_2022_QST}.
However, in our case, the value of $\xi$ is always positive. Thus, the negative contribution of the nonlinear terms in Eq.~(\ref{eq:effHamiltonian_b}) modifies the critical point such that the effective strength is getting closer to the critical point, i.e. $\Lambda_c(\xi{>} 0)<\Lambda_c(\xi{=}0)$.

The finite $\xi$-induced enhancement for steady state QFI can also be understood via the quantum coherence~\cite{Plenio_PRL_2014_coherence, Plenio_PRL_2017_resource}. One intuitively expects that the dominant resource of QFI here is the coherence. To prove this quantitatively we employ the $l_1$ norm of coherence $C_{l_1}=\tfrac{1}{2}\sum_{j\neq i}|\rho_{i,j}|$ as a measure of coherence and compute it for the steady-state of our effective system~\cite{Plenio_PRL_2014_coherence}.
For $\xi=0$ rapid growth of $C_{l_1}$ as the system approaches the critical point sets a direct relation with the critically enhanced QFI, see Fig.~\ref{fig:sensing}(d).
The nonlinear contribution with strength $\xi$ to the system generates a superposition of the squeezed number states $\{ |n\rangle_s \}$~\cite{Devoret_nature_2020_catstate}.
This leads to a further layer of superposition in terms of the Fock basis $\{|n\rangle\}$ that increases the coherence in the system.
The results for coherence have been shown in Fig.~\ref{fig:sensing}(d), which numerically confirms our claim.

\subsection{Finite Temperature}

\begin{figure}
    \centering
    \includegraphics[width=\columnwidth]{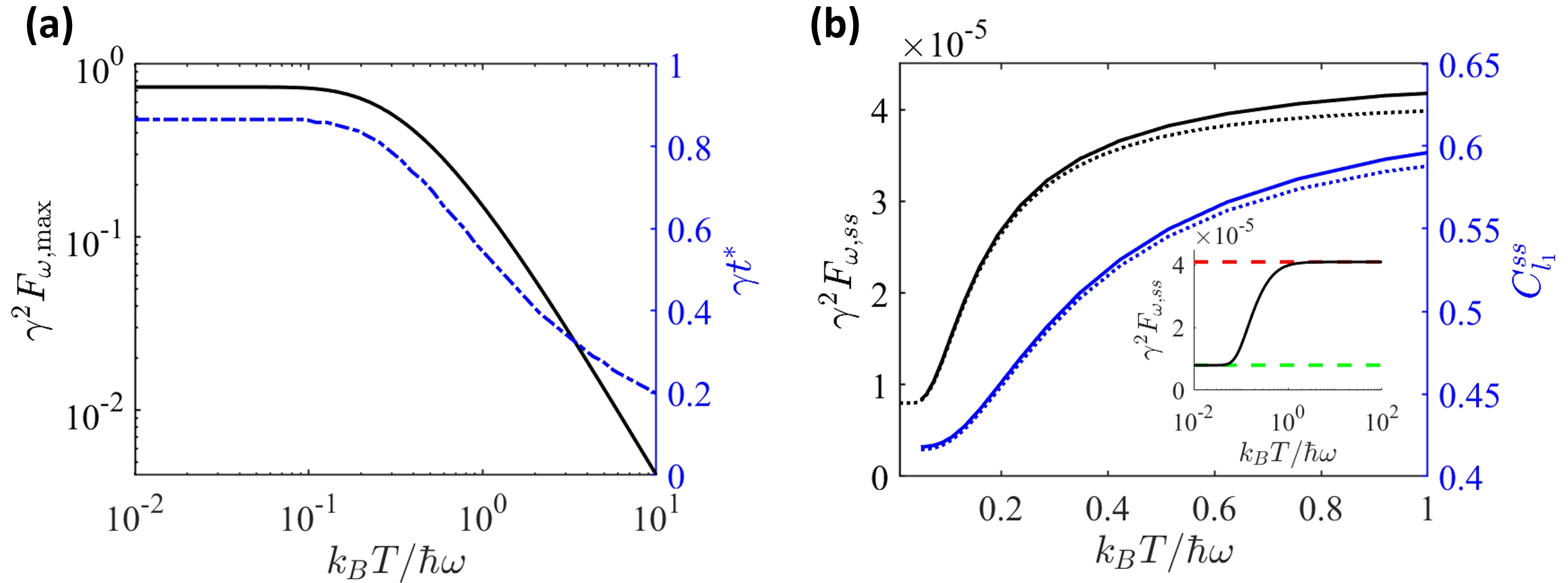}
    \caption{QFI for finite temperature. (a) The maximal QFI and the corresponding optimal $t^{*}$ vs temperature $T$ when $\xi=0$. (b) The steady states QFI for finite temperatures. The dashed-dot lines correspond to the steady state result for $\xi=0$ while the solid lines indicate the steady states with $\xi=5\times 10^{-5}\omega$. Inset: steady-state QFI for $\xi=0$, the red and the green dashed lines correspond to the high and the low temperature limits, respectively, see Eq.~\eqref{eq:ss_Qfi_T} and the discussion below. In this plot, we choose $\Lambda=0.8\Lambda_c$.}
    \label{fig:sensingT}
\end{figure}

Due to the limited cooling efficiencies in realistic quantum systems, it is necessary to analyze the frequency estimation for arbitrary effective temperatures. The thermal fluctuation for $T{>}0$ drives the system to the classical regime when $k_B T{\gg}\hbar\Delta_{kj}$ and the quantum criticality vanishes.
Hence, the phenomenological prediction is the suppression of the QFI close to the critical point. The decreasing $F_{\omega,\text{max}}$ shown in Fig.~\ref{fig:sensingT}(a) confirms the destruction of critical QFI in the dynamical case.
However, surprisingly, the steady-state QFI at finite $T$ indicates that the thermal fluctuations could induce an enhancement for the sensitivity.
Indeed, it is noticed that for $\xi=0$, the system eventually approaches a squeezed thermal state.
The steady-state QFI for finite $T$ thus is calculated with the exact form
\begin{equation}
    F_{\omega,ss} = \frac{(\partial_{\omega} \Delta)^2 (\partial_{\Delta}\bar{n})^2}{\bar{n}(\bar{n}+1)} + \frac{4\Lambda^2}{\Delta^4}\frac{2(2\bar{n}+1)^2}{2\bar{n}^2 +2\bar{n}+1}.
\label{eq:ss_Qfi_T}
\end{equation}
The first part is the conventional contribution of QFI at thermal equilibrium, while the second part indicates the influence of squeezing.
In the low-temperature limit $\bar{n}{\rightarrow} 0$ the thermal contribution fades away as $(\partial_{\Delta}\bar{n})^2/\bar{n}(\bar{n}+1){\sim} 0$ and thus $F_{\omega,ss}(T{\to} 0)=8\Lambda^2/\Delta^4$, while for high-temperatures $\bar{n}{\gg} 1$, we find $F_{\omega,ss}(T{\to} \infty)\approx(\omega^2+16\Lambda^2)/\Delta^4$.
In Fig.~\ref{fig:sensingT}(b) the variation of steady-state QFI with temperature and its asymptotical behaviors are shown.
It is worth mentioning that the steady-state QFI for a squeezed thermal state scales as $F_{\omega,ss}{\propto}\Delta^{-4}$ regardless of the temperature regime. This surpasses the scaling behavior of QFI for Gibbs states~\cite{Absiuso_PRL_2025_metrology}. The enhancement of $F_{\omega,ss}$ with temperature can be explained through the coherence of the system at its steady-state. Indeed, the squeezed thermal states have larger coherence compared to the thermal states. Therefore, the larger effective temperature $T$ mixes more squeezed Fock states to enhance quantum coherence $C^{ss}_{l_1}$, the blue lines in Fig.~\ref{fig:sensingT}(b). The finite nonlinearity $\xi$ generates additional coherence, similar to the zero temperature case.

\section{Conclusion}\label{sec:conclusion}

We have proposed and studied a hybrid quantum device based on the graphene Josephson junctions. By considering the motional DOF of the monolayer graphene membrane we engineer a tunable nonlinear coupling between the vibrational mode and the superconducting circuit. 
The parametric process between the circuit and the vibrational mode can be used to effectively cool down the vibrational mode close to the ground-state level. To address the potential application of our proposed system, we have discussed the generation of mechanical cat states and the critical role of quantum metrology in estimating the mechanical oscillating frequency. The strong, effective nonlinearity and the small dissipation rate ensure the fast generation and manipulation of mechanical cat states. For critical quantum sensing, we also argue that the coherence produced in the system serves as a resource for enhancing QFI.
Due to the intrinsic nonlinearity of the mechanical object and the flexible control for the interacting strength, our proposed setup and protocols provide potential applications in mechanical quantum information processing.

\begin{appendix}
\section{Graphene Josephson Junction}\label{ap:Andreev}
In this section, we briefly review the detailed form of the Graphene Josephson junctions, which have been studied by Titov and Beenakker~\cite{Beenakker_PRB_2006}. By considering the low energy range $\varepsilon < \Delta_0$, where the energy of the Andreev bound state is below the 
 excitation gap for the superconductor islands $\Delta_0$, the Andreev reflection occurs at each Graphene-superconductor interface, and the wave vector $k_y$ parallel to the interface is conserved. For a finite width of graphene, the wave vector parallel to the interface can be quantized, for which $k_y = q_n = (n+1/2)\pi/W$ by taking the "infinite mass" boundary conditions at $y=0,W$. 

The single Andreev bound state energy per mode is
\begin{equation}
\varepsilon_n = \Delta_0 \sqrt{1-\tau_n \sin^2(\phi/2)},
\end{equation}
where $\phi$ is the superconducting phase difference, and
\begin{equation}
\tau_n = \frac{k^2_n}{k^2_n \cos^2(k_n L) + (\mu/\hbar v_F)^2 \sin^2(k_n L)},
\end{equation}
the transmission probability for each Andreev mode, with $k_n = \sqrt{(\mu/\hbar v_F)^2 - q^2_n}$. The Josephson current at zero temperature is given by
\begin{equation}
    I = -\frac{4e}{\hbar}\frac{d}{d\phi}\int d\varepsilon \sum_{n=0}^{\infty}\rho_n(\varepsilon,\phi)\varepsilon.
\end{equation}
Here $\rho_n(\varepsilon,\phi)$ is the density of states. For the dispersionless case, $\rho_n(\varepsilon,\phi) = \delta(\varepsilon-\varepsilon_n(\phi))$. The voltage across the junction is time-dependent to the phase difference, according to
\begin{equation}
V = \frac{d\phi}{dt}\frac{\Phi_0}{2\pi}.
\end{equation}
Here $\Phi_0 = 2\pi \hbar/4e$ is the flux quantum, where the factor 4 indicates the twofold spin and valley degeneracies for graphene. Therefore, the Josephson potential energy is obtained according to
\begin{equation}
    \begin{split}
    H_{\text{GJJ}} &= \int dt V I \\
    &= -\int \frac{\hbar}{4e}\frac{d\phi}{dt}\frac{4e}{\hbar} \frac{d}{d\phi}\sum_{n=0}^{\infty} \varepsilon_n(\phi) \\
    &= -\Delta_0 \sum_{n=0}^{\infty}\sqrt{1-\tau_n \sin^2(\phi/2)}.
    \end{split}
\end{equation}

\begin{widetext}

\section{Effective Master Equation}\label{ap:effmastereq}
\subsection{zero $\xi$}
For $\xi \rightarrow 0$, the effective Hamiltonian can be exactly diagonalized under the squeezing transformation, as indicated in Eq.~\eqref{eq:effHamiltonian_sqz}. Therefore, starting from the reservoir Hamiltonian $H_R=\sum_j \omega_j r^{\dagger}_j r_j$ and the system-reservoir interaction $H_{SR}=\sum_j \gamma^*_j r^{\dagger}_j b + \gamma_j r_j b^{\dagger} = b R^{\dagger}+b^{\dagger}R$, where we denote $R=\sum_j \gamma_j r_j$ in short. After the squeezing transformation, the total system-reservoir Hamiltonian has the form
\begin{equation}
    H_{SR} = (b\cosh r +b^{\dagger}\sinh r)R^{\dagger} + (b^{\dagger}\cosh r + b\sinh r)R.
\end{equation}
Transforming into the interacting picture and following the standard Born-Markovian approximation, we have
\begin{equation}
    \begin{split}
        \dot{\rho} &= -\int_0^{\infty} d\tau \text{Tr}_R \left\{ [H_{SR}(t),[H_{SR}(t-\tau),\rho(t)]] \right\} \\
        &= -\int_0^{\infty} d\tau \left\{ [ (\cosh^2(r) bb^{\dagger} e^{-i\Delta \tau} + \sinh^2(r) b^{\dagger}be^{i\Delta \tau})\rho(t) -\cosh^2(r) b^{\dagger}\rho(t)be^{-i\Delta \tau} \right.\\
        &\left.- \sinh^2(r) b\rho(t)b^{\dagger}e^{i\Delta \tau} ] \langle R^{\dagger}(t)R(t-\tau)\rangle_R + \text{h.c.} \right.\\
        &\left. + [(\cosh^2(r) b^{\dagger}b e^{i\Delta \tau} + \sinh^2(r) b b^{\dagger}e^{-i\Delta \tau})\rho(t)  - \cosh^2(r) b\rho(t)b^{\dagger}e^{i\Delta \tau} \right.\\
        &\left. -\sinh^2(r) b^{\dagger}\rho(t)be^{-i\Delta \tau}]\langle R(t)R^{\dagger}(t-\tau)\rangle_R +\text{h.c.} \right\}.
    \end{split}\label{eq:der_meq1}
\end{equation}
Without loss of generality,  the reservoir correlating terms can be approximated as
\begin{equation}
    \begin{split}
        \langle R^{\dagger}(t) R(t-\tau) \rangle_R &\sim \int_0^{\infty} d\omega g(\omega)|\gamma(\omega)|^2 e^{i\omega \tau} \bar{n}(\omega,T),\\
        \langle R(t)R^{\dagger}(t-\tau) \rangle_R &\sim \int_0^{\infty} g(\omega)|\gamma(\omega)|^2e^{-i\omega \tau} [\bar{n}(\omega,T)+1].
    \end{split}\label{eq:der_meq2}
\end{equation}
Let $\gamma=\pi g(\Delta)|\gamma(\Delta)|^2$ and transforming back to the Schrodinger picture, we have the effective master equation
\begin{equation}
    \dot{\rho} = -i[H_{\text{eff}},\rho] + \Gamma (\bar{n}(\Delta,T)+1) D_{b}[\rho] + \Gamma \bar{n}(\Delta,T)D_{b^{\dagger}}[\rho].\label{eq:meq_gauss}
\end{equation}
Where we define $\Gamma=\gamma\cosh^2{r}$ as the effective dissipation. The steady-state for the above master equation has the form
\begin{equation}
\begin{split}
\rho_{ss} &= \sum_{n=0}^{\infty} P_n |n\rangle_s \langle n|,\\
    P_n &= \frac{1}{1+\bar{n}} \left( \frac{\bar{n}}{1+\bar{n}} \right)^n ,
    \end{split}\label{eq:squFock}
\end{equation}
This state is the squeezed thermal state, where $|n \rangle_s = S(r)|n\rangle$ indicates squeezed Fock states, $\bar{n}$ denotes $\bar{n}(\Delta,T)$ in short. The steady-state QFI for the squeezed thermal states (Eq.~\eqref{eq:ss_Qfi_T}) can then be obtained by substituting Eq.~(\ref{eq:squFock}) into the definition of QFI through the logarithmic derivative $F_o=2\sum_{j,k}|\langle\psi_j|\partial_o \rho|\psi_k\rangle|^2/(\lambda_j +\lambda_k)$.

\subsection{Finite $\xi$}

For finite nonlinearity, when the system is getting closer to the critical point, the effective anharmonicity is critically enhanced. In that case, the environment-induced transitions largely depend on the eigenstates, which can not be perturbatively expanded under either the original Fock basis ${|n\rangle}$ or the squeezed Fock basis. To describe the dissipative dynamics between the eigenbasis, we derive the effective master equation under the eigenbasis $H_{\text{eff}}|k\rangle_s=E_k |k\rangle_s$ for the effective system described by Eq.~ (8) in the main text. Then $b/b^{\dagger}$ in the eigenbasis can be denoted as
\begin{equation}
    \begin{split}
        b &\rightarrow \sum_{j,k} \prescript{}{s}\langle j|b|k\rangle_s |j\rangle_s \langle k|\equiv s,\\
        b^{\dagger} &\rightarrow \sum_{j,k} \prescript{}{s}\langle j|b^{\dagger}|k\rangle_s |j\rangle_s \langle k|\equiv s^{\dagger},
    \end{split}
\end{equation}
Here, we assume the reservoir has the Ohmic spectral density in the low-frequency limit. Therefore, the Born-Markovian approximation remains valid even in the vicinity of the critical regime. Following the standard derivation (Eq.~\eqref{eq:der_meq1}) and consider the same reservoir correlation (Eq.~\eqref{eq:der_meq2}), we have
\begin{equation}
\begin{split}
\dot{\rho} &= -\int_0^{\infty} d\tau \text{Tr}_R \left\{ [H_{SR}(t),[H_{SR}(t-\tau),\rho(t)]] \right\} \\
& = -\int_0^{\infty} d\tau \left\{ [ \sum_{j,k} \prescript{}{s}\langle j|b|k \rangle_s\langle k|b^{\dagger} |j\rangle_s|j \rangle_s \langle k|\times|k\rangle_s \langle j|e^{-i\Delta_{kj}\tau}\rho(t) \right.\\
&\left. - \sum_{j,k} \prescript{}{s}\langle j|b^{\dagger}|k \rangle_s\langle k|b|j\rangle_s |j\rangle_s \langle k|\rho(t)|k\rangle_s \langle j|e^{-i\Delta_{jk}\tau} ] \langle R^{\dagger}(t) R(t-\tau)\rangle_R +\text{h.c.} \right.\\
&\left. + [\sum_{j,k} \prescript{}{s}\langle j|b^{\dagger}|k \rangle_s\langle k|b|j\rangle_s| j \rangle_s \langle k|\times|k\rangle_s \langle j|e^{-i\Delta_{kj}\tau}\rho(t) \right.\\
&\left. - \sum_{j,k} \prescript{}{s}\langle j|b^{\dagger}|k \rangle_s\langle k|b|j\rangle_s |j\rangle_s \langle k|\rho(t)|k\rangle_s \langle j|e^{-i\Delta_{jk}\tau}]\langle R(t)R^{\dagger}(t-\tau)\rangle_R +\text{h.c.} \right\}.
\end{split}
\end{equation}
In the above derivation, all the terms $e^{i(\Delta_{jk}+\Delta_{mn})t}$ that $\Delta_{jk}+\Delta_{mn}\neq 0 $ have been dropped as counter-rotating terms. 
Therefore, the master equation for the effective system, as presented in Eq.~\eqref{eq:effmastereq}, is obtained.

\end{widetext}

\end{appendix}

\bibliography{ref_GbasedQD}

\end{document}